# The pseudogap and doping dependent magnetic properties of $La_{2-x}Sr_xCu_{1-y}Zn_yO_4$

R. S. Islam[1,2], J. R. Cooper[1], J. W. Loram[1], and S. H. Naqib[1,2*]

[1]Department of Physics, University of Cambridge, J. J. Thomson Avenue, Cambridge CB3 0HE, United Kingdom

[2]Department of Physics, University of Rajshahi, Raj - 6205, Bangladesh

The effects of planar hole content, $p$ (= x), on the static magnetic susceptibility, $\chi(T)$, of polycrystalline $La_{2-x}Sr_xCu_{1-y}Zn_yO_4$ compounds were investigated over a wide range of Sr (x) and Zn (y) contents. The magnetic behavior caused by Zn was found to depend strongly on the hole content. The apparent magnetic moment induced by Zn was larger in underdoped $La_{2-x}Sr_xCu_{1-y}Zn_yO_4$, decreased quite sharply around $p \approx 0.19$, and did not change much for further overdoping. This is interpreted in terms of the effect of the pseudogap on the Zn-induced magnetic behavior, as there is growing evidence that the pseudogap vanishes quite abruptly at $p \approx 0.19 \pm 0.01$. From a detailed analysis of $\chi(T)$ data the Zn-induced magnetic contribution was found to be rather complex and showed non-Curie-like features over a wide range of temperature. The observed behavior was scrutinized in terms of two scenarios (a) that of independent localized-moments and (b) low energy quasiparticle resonances associated with each Zn atom. Our study points towards the latter scenario and more generally suggests that there is a re-distribution of quasiparticle spectral weight due to Zn substitution, the features of which are greatly influenced by the presence and magnitude of the pseudogap.

*PACS numbers:* 74.72.Dn, 74.25.Ha, 74.25.-Dw, 74.62.Dh

*Keywords*: La-based cuprates, Impurity effects, Magnetic susceptibility, Pseudogap

## I. INTRODUCTION

Cuprate superconductivity arises from strongly interacting charge carriers in the $CuO_2$ planes. Substituting a single impurity for a copper atom perturbs the surrounding electronic environment and can therefore be used as a probe for understanding the physics of high-temperature superconductors (HTS). Experiments on Zn-substituted cuprates provide valuable information on the electronic properties of the host materials, in both the normal and the superconducting (SC) states. $Zn^{2+}$ is a non-magnetic ion owing to its $3d^{10}$ (spin $s = 0$) electronic configuration, and is a very strong destroyer of $d$-wave superconductivity. It has been shown in early years [1] that Zn causes a rapid suppression of $T_c$, ~ -10K/%Zn in the $CuO_2$ plane. Zn is believed to act as a unitary scatterer [2,3] in suppressing the superconductivity of the host material, without substantially altering the planar hole concentration, $p$ [4]. Electron spin [5] and nuclear magnetic resonance [6,7] experiments have indicated that non-magnetic Zn gives rise to a local magnetic moment-like feature on the four nearest neighbor Cu sites in the $CuO_2$ planes. This apparent magnetic behavior is also evident from the appearance of a 'Curie-like' term in the bulk magnetic susceptibility of the Zn-doped samples [3,8-10]. The effect is thought [3] to arise from the suppression of local short-range antiferromagnetic (AF) correlations as a result of replacing ($s = 1/2$) $Cu^{2+}$ with ($s = 0$) $Zn^{2+}$. Scanning tunneling microscopy (STM) experiments on Zn-substituted $Bi_2Sr_2CaCu_2O_{8+\delta}$ single crystals [11] found intense low-energy quasiparticle (QP) scattering resonances at the Zn sites, coincident with the strong suppression of superconductivity within ~ 15Å of the scattering sites. Complementary results were obtained earlier by Loram *et al.* from their specific heat measurements [12], namely a strong increase in the low temperature ($T \rightarrow 0K$) electronic specific heat coefficient, $\gamma(0)$, with Zn. This implies that Zn is a strong pair breaker that induces a significant residual density of states (DOS) near the Fermi level in the superconducting state. There are also strong arguments in favor of a transfer of spectral weight from high energy to low energy in Zn substituted compounds [13] that is supported by some inelastic neutron scattering experiments [14]. This is interesting because it opens up the possibility of an alternative scenario for explaining the magnetic properties of Zn doped cuprates, in terms of a low energy resonance affecting the spin susceptibility, involving a re-distribution of QP spectral weight rather than the formation of independent localized magnetic moments. A detailed investigation is still lacking.

In this paper, we report some systematic studies of the effects of planar hole content on the static magnetic susceptibility, $\chi(T)$, of $La_{2-x}Sr_xCu_{1-y}Zn_yO_4$ (Zn-LSCO) sintered samples over a wide range of Sr (x) and Zn (y) contents. From the analysis of the $\chi(T)$ data, we have found the Zn-induced Curie-like increase in magnetic susceptibility to be strongly dependent on the hole content and to correlate closely with the presence of the pseudogap (PG). The effect of Zn was also found to be strongly $T$-dependent, however the behavior does not follow a simple Curie-Weiss $T$-dependence over the whole temperature range. Instead the temperature dependence is rather complex and we argue that the localized moment scenario, supplemented by the Kondo effect does not offer a complete explanation. Our analysis shows that the more likely scenario is a Zn induced re-distribution of QP spectral weight that depends strongly on the existence and size of the PG.

## II. EXPERIMENTAL DETAILS AND RESULTS

Polycrystalline single-phase sintered samples of $La_{2-x}Sr_xCu_{1-y}Zn_yO_4$ were synthesized by standard solid-state reaction methods using high-purity (> 99.99%) starting materials. Samples were characterized by X-ray powder diffraction, room-temperature thermopower, and AC susceptibility measurements. Details of sample preparation and characterization can be found elsewhere [15]. $\chi(T)$ measurements were made using a model 1822 Quantum Design SQUID magnetometer. During the measurement the sample was mounted between two quartz tubes of similar dimensions that were attached to the sample rod. The tubes were properly cleaned before each measurement to avoid contamination by any magnetic particles. Data were collected following a predefined sequence, usually in the range of 5K to 400K, and for different values of the field so as to check the linear field dependence. A scan length of 6cm was used. The background signal was due to the absence of quartz at the position of the sample and varied linearly with the separation between the two quartz tubes up to a separation of approximately 6mm, the length of the samples. Hence the shape of the SQUID response curve was the same for the sample and for the empty sample holder. This meant that the magnetic moment of the quartz tubes could be subtracted from the data to obtain the magnetic moment of the sample, from which the susceptibility was then obtained. We have investigated five Sr concentrations, two underdoped (9% and 15% Sr) and three overdoped (19%, 22%, and 27% Sr). For each Sr content we have measured six Zn contents (0, 0.5, 1.0, 1.5, 2.0, and 2.4% Zn) chosen to be within the solubility limit.

We attribute anomalous susceptibility data at around 60K for some of the samples to the presence of absorbed oxygen.

The $\chi(T)$ data are shown in Figs. 1. The $\chi(T)$ plots show that (i) a Curie-like term appears at low-temperature and that this increases with Zn concentration, (ii) this Curie-like term decreases with increasing Sr content (i.e. with increasing hole doping), and (iii) for the samples with x > 0.19, this term is similar in magnitude to that for x = 0.19. Therefore, it is apparent that the Zn-induced increase in $\chi(T)$ is larger for underdoped $La_{2-x}Sr_xCu_{1-y}Zn_yO_4$.

## III. DATA ANALYSIS

In the present study we wish to determine the added contribution, $\chi_{Zn}(T)$, due to Zn substitution. Since systematic errors may be present in any given sample, for example errors in the Sr content, as well as SQUID measurement errors near 60K arising from absorbed oxygen, we wish to avoid placing undue weight on the Zn-free sample. We therefore apply a linear fit at each experimental temperature $T$ of the form

$$\chi(T) = a_0(T) + a_1(T)y \qquad (1)$$

where $a_0(T)$ and $a_1(T)$ are the coefficients determined from least-square fits to the $\chi(T)$ data for a given value of Sr concentration (x). Evidently $a_0(T)$ gives the fitted contribution from the Zn-free sample (y = 0) and $a_1(T)$ is the increment per %Zn. The justification for using a linear fit comes from the experimental data, but is expected on general grounds (above the superconducting region) for the low levels of substitution used, up to 2.4% Zn/Cu. In Fig. 2 we show $\chi(T)$ at several different fixed temperatures $T_0$, versus Zn content (y) for 15% Sr doped LSCO, and this clearly shows the validity of the linear approximation. The results of least-square fits, $\chi_{fit}(T)$, are shown in Fig. 3 where the lower temperature limit has been chosen to avoid the region of significant diamagnetic SC fluctuations. This is easily recognized as the temperature below which the variance and residuals sharply increase. Comparing $\chi_{fit}(T)$ with the raw data in Fig. 1 confirms the quality of the fits. We have also analyzed the error $\Delta\chi(T) = [\chi(T) - \chi_{fit}(T)]$ for the respective compounds. It can be seen from Fig. 4 that for most of the compounds $\Delta\chi(T)$ is less than 2% of the experimental $\chi(T)$ at 400K. From the *T*-dependences of $\Delta\chi(T)$ we conclude that the systematic errors result mainly from small differences in Sr content at the level of ± 0.5%.

The coefficient $a_0$, representing the fitted values for the Zn-free samples, is shown in Fig. 5a. The absence of a pseudogap for *p* > 0.19 is evident from this figure. From previous analyses of magnetic susceptibility and specific heat data [15,16] we have extracted the PG energy scale *T**, in degrees K, for the LSCO compounds. *T** = 500K for *p* = 0.09 and *T** = 290K for *p* = 0.15 [15,16]. These values correspond approximately to the temperature of the shoulder in $\chi$. Interestingly they are both a factor of 1.3 larger than the crossing temperatures ($T_{cs}$) denoted by arrows in Figs. 3a and 3b. Next, the *T*-dependence of the coefficient $a_1$ is plotted in Fig. 5b for different values of Sr content, and shows the change in magnetic susceptibility per %Zn. $a_1(T)$ shows a Curie-like increase at low temperatures for all samples, with a magnitude that decreases strongly with hole doping. For *p* > 0.19 the Curie-like term is small and almost p-independent.

## IV. DISCUSSION

When conventional localized magnetic moments with a low concentration y are introduced into a normal metal the contribution to the susceptibility can usually be expressed as the sum of two independent contributions

$$\Delta\chi(y, T)/y \equiv a_1 = C/(T+\theta) + \Delta\chi_0 \qquad (2)$$

where the Curie-Weiss (CW) term $C/(T+\theta)$ reflects the local response of the magnetic moments and the 'dilution' term $\Delta\chi_0$ expresses any change in the host susceptibility and is usually taken to be temperature independent. The Curie temperature $\theta$ reflects the characteristic energy of the magnetic excitations due, for example, to a Kondo effect, direct-exchange, or RKKY interactions. It is clear from Fig. 5c showing plots of $a_1(T)T$ versus $T$ that the underdoped samples exhibit a more complex behavior than suggested by Eq. 2, and can only be described by this expression if $\Delta\chi_0$ is allowed to be strongly $T$-dependent, as discussed below. Instead we believe that our data is better described in terms of a re-distribution of spectral weight from high to low energies within a single excitation spectrum.

In order to clarify this we first discuss the $a_1(T)T$ plots in terms of Eq. 2. At low $T$, $a_1(T)T$ is relatively constant with no obvious linear term that could be attributed to $\Delta\chi_0 T$. Thus the 'dilution' term $\Delta\chi_0$ appears to be very small at low temperatures, within the range $\Delta\chi_{0LT} = 0 \pm 0.007 \times 10^{-4}$ emu/mol%Zn. Also, according to Eq. 2, $a_1 T$ should fall towards zero as $T$ falls below $\theta$. The experimental decrease is at most very weak, and we conclude that $\theta$ is substantially smaller than 40K (our low-$T$ cut-off due to the onset of superconductivity). Approximate estimates of $\theta$ can be obtained for the more heavily Zn doped samples with low or zero $T_c$ by subtracting from $\chi$ a linear extrapolation of the Zn free susceptibility from above $T_c$ to $T = 0$K. This yields values of $\theta \sim (15 \pm 5)$K for the underdoped samples. For the overdoped samples the smaller CW term, uncertainties in the extrapolation of the Zn-free susceptibility, sensitivity to errors in Sr content and the presence of SC fluctuations even for heavy Zn doping preclude a reliable estimate of $\theta$. The Curie constants $C_{LT}$ given by the approximately constant values of $a_1 T$ in the low temperature region are shown in Fig. 6, and decrease strongly up to 19% Sr but show little change at higher doping. From $C_{LT}$ we obtain effective magnetic moments $p_{eff}$ = 0.98, 0.80, 0.67, 0.54 and 0.49 $\mu_B$/Zn, for x = 0.09, 0.15, 0.19, 0.22, and 0.27 respectively, consistent with values found in earlier studies [3,17-19]. Above a certain temperature ($\sim 100 - 150$ K) which for x = 0.09 and 0.15 is $\sim T^*/3$, $a_1(T)T$ develops a negative slope that can be attributed to an increased negative 'dilution' term $\Delta\chi_0$, and which is large for the two underdoped samples with 9% and 15% Sr. High temperature values of the Curie constant $C_{HT}$ and 'dilution' term $\Delta\chi_{0,HT}$ are also plotted in Fig. 6, these quantities having been determined from the slope and intercept of linear fits of $a_1$ vs. $1/T$ in the temperature region $\sim 200$K to 400K (see Fig. 5d). It is significant that the $p$-dependences of $C_{HT}$ and $\Delta\chi_{0,HT}$ are correlated, with both showing a sharp increase below 19% Sr where the pseudogap opens. Note that the values of $C_{HT}$ are almost a factor two higher than $C_{LT}$ for $p \leq 0.15$ but are almost identical to $C_{LT}$ for $p \geq 0.19$. For $p \geq 0.22$ the values of $\Delta\chi_{0,HT}$ are small and comparable (per %Zn) to a reduction of ~ 1% in the susceptibility of the Zn-free material at 400K, as shown by the dashed line in Fig. 6. In this doping region it is therefore reasonable to ascribe $\Delta\chi_0$ to a "dilution" effect. For the two underdoped samples $\Delta\chi_0$ is a factor six larger, and is, we believe, too large to be interpreted in this way (note that a 4x larger dilution effect was assumed in Ref. [6]). In summary, interpreting our data in terms of the two component expression Eq. 2 is unsatisfactory, particularly for the underdoped samples containing 9% and 15% Sr. Values of the Curie constant $C$ and "dilution" term $\Delta\chi_0$ differ strongly when estimated at low and high temperatures and at high temperatures the two parameters are clearly not independent. The low temperature value $C_{LT}$ is undoubtedly the more reliable since the CW term dominates the susceptibility in this temperature region. The high temperature value $C_{HT}$, estimated from the high temperature slope of $a_1$ vs. $1/T$ (Fig. 5d) assuming a $T$-independent $\Delta\chi_0$ cannot be trusted if $\Delta\chi_0$ is in fact strongly $T$-dependent.

A revealing presentation of the data which does not rely on any model is given by plots of $a_0^*$ and $a_1^*$ vs. $T$ shown in Fig. 7. We define $\chi^* = d(\chi T)/dT$, with corresponding expressions $a_0^*$ and $a_1^*$ for the susceptibility components $a_0$ and $a_1$. Whilst $\chi$ reflects the DOS of spin excitations averaged over a thermal energy window extending from zero to ~ $2k_BT$, it can be shown that $\chi^*$ reflects the DOS averaged over an energy region ~ $2k_BT$ ± $k_BT$. In the particular case of a Fermi system with Fermi energy $E_F$, $\chi$ is sensitive to states between $E_F$ and ~ $E_F$ ± $2k_BT$, while $\chi^*$ is sensitive to states between ~ $E_F$ ± $k_BT$ and ~ $E_F$ ± $3k_BT$. Hence $\chi^*$ excludes the zero and low energy excitations that make a large contribution to $\chi$ at all temperatures and provides a sharper representation of the spin DOS. For example the Curie-Weiss term $C/(T+\theta)$ which dominates the present $\chi$ data over almost the entire $T$-range becomes a much narrower function $C\theta/(T+\theta)^2$ in $\chi^*$. This latter function reflects more closely the energy range ~ $k_B\theta$ of the underlying excitations. If $\chi$ is $T$-independent, then $\chi^* = \chi$.

The reduction in width of the pseudogap with increasing p and its absence for x > 19% Sr is evident in the plots of $a_0^*(T)$ for the Zn-free material (Fig. 7a). The increment $a_1^*$ per % Zn (Fig. 7b) also shows strikingly different behavior below 19% Sr. $a_1^*(T)$ for the 19%, 22% and 27% Sr samples is small, negative and *T*-independent, with magnitude similar to the values of $\Delta\chi_{0,HT}$ shown in Fig. 6. By contrast, $a_1^*$ for the 9% and 15% Sr samples has large and broad minima at temperatures ~ 400K and 230K respectively, comparable with the temperatures where $a_0^*(T)$ flattens out into a broad peak. The presence of these minima is consistent with the striking difference between the large high-*T* and small low-*T* values of $\Delta\chi_0$ for these two samples, obtained from fits to the two component expression Eq. 2.

The dominance of SC correlations below 40K and the low value $\theta \sim 15$K tend to obscure evidence for a low temperature upturn, associated with a finite value of $\theta$, in the plots of $a_1^*$ shown in Fig. 5b. (Recall that $a_1^*$ is not affected by a pure Curie term with $\theta = 0$K). However we can estimate $a_1^*$ at lower temperatures for heavily Zn doped samples with low or zero $T_c$ from $a_1^* \sim (\chi^*(y, T) - \chi^*_{extrap}(0, T))/y$ where $\chi^*_{extrap}(0, T)$ is a linear extrapolation of the Zn-free data from above $T_c$ to $T = 0$K. This is shown in Fig. 8 for the 15% Sr sample containing 2.0 and 2.4% Zn. The upturn per %Zn below 60K is independent of y, confirming the absence of SC correlations, and a fit to $C\theta/(T+\theta)^2$ yields $\theta \sim 15$K and a Curie constant $C$ close to the value $C_{LT}$ shown in Fig. 6. The positive low temperature peak together with the negative area under the $a_1^*$ curve at higher temperatures, strongly indicate that the growth of the low energy peak is at the expense of spectral weight from higher energies in the normal state spectrum. This is also demonstrated in the inset to Fig. 8, where we show $\chi^*$ for 15% Sr containing 0% Zn (with a linear extrapolation below 60K) and raw $\chi^*$ data for 2.4% Zn. It is evident from this plot that on doping with Zn weight is lost in the normal state spectrum predominantly from the region of the peak in $\chi^*$ for the Zn-free material. For the 9% Sr sample a similar coincidence of a negative peak in $a_1^*$ and the maximum in $a_0^*$ would be consistent with the data (Figs. 7a and 7b), though here both features occur at or a little above 400K.

If we accept this circumstantial evidence that the low energy peak in Zn doped samples is derived from higher energy states in the normal state spectrum of the Zn-free host, then clearly these states must have the same character (*i.e.,* spin or QP). The strongly temperature dependent susceptibilities of the pure LSCO system (see *e.g.,* Figs. 1 and 5a) was originally attributed [20] to short range antiferromagnetic correlations of the Cu spins. In this interpretation, the broad susceptibility maxima occurring at temperatures $T_{peak}$ that decrease with hole doping (Sr content) reflect the mean energy of the short range spin correlations. In this scenario it would be natural to ascribe the Zn-induced Curie term to a purely spin effect due to neighbouring Cu [3,6,7]. We question this interpretation for two reasons. Firstly, it seems improbable that the characteristic energy of the Cu spins neighbouring the Zn impurity (~ $k_B\theta$ ~ 15K) should be more than an order of magnitude lower than the interaction energies of more distant Cu spins (~ $k_B T_{peak}$ ~ 300 – 500K) since these latter spins are already heavily disordered. The situation is even worse if $\theta$ reflects a Kondo temperature [21,22] since then the exchange energies of the neighbouring Cu spins must be substantially lower than $k_B\theta$, otherwise the Kondo effect would be suppressed. Secondly, it is a clear prediction of the short range AF order scenario that the magnitude at the peak $\chi(T_{peak}) \propto 1/T_{peak}$, since the number of Cu spins is doping independent. This is not observed (see Fig. 1 or 5a), casting strong doubt that the normal state spectrum is dominated by spin excitations, and that the low energy gap is a spin gap. Instead we have argued [16] from the close correspondence between the susceptibility and the electronic specific heat ($\chi \sim R_0 S/T$ where $S$ is the electronic entropy and $R_0$ is the free electron Wilson ratio), that both quantities are dominated by the QP spectrum and that the temperature dependences reflect the energy-dependence of the QP DOS, in particular the presence of a PG for Sr content $x \leq 0.19$. This conclusion has been confirmed by ARPES data on LSCO [23]. We therefore conclude [13] that the Zn-induced Curie-like magnetism results from a narrow QP resonance located near $E_F$, and that these resonant states are drawn predominantly from the shoulder region of the V-shaped PG which is, in turn, responsible for the shoulder in the susceptibility. A narrow QP resonance induced by unitary impurity scattering in the presence of a gap at $E_F$ has been proposed theoretically. This was originally proposed for a $d$-wave superconductor [24] and later applied [25] to STM studies [11] of low energy resonant states induced by Zn impurities in the SC gap of Bi2212. Related work [26] showed that the origin of the gap at $E_F$ is immaterial, and that impurity-induced resonances would also be expected within a PG at $E_F$. This interpretation readily explains (a) the large Curie term accompanied by a substantial suppression of spectral weight at higher energies due to Zn doping in samples with a pseudogap, (b) its abrupt decrease as the pseudogap closes at 19% Sr, and (c) the rather weak effect of Zn doping on the susceptibility of more heavily doped samples where the pseudogap is absent.

## V. CONCLUSIONS

In summary, we have studied the static magnetic susceptibility of $La_{2-x}Sr_xCu_{1-y}Zn_yO_4$ over a wide range of hole concentration and Zn contents. Non-magnetic Zn doping is found to induce a Curie-Weiss term in the susceptibility which is large in underdoped materials with a pseudogap ($x < 0.19$) but is rather small and doping-independent in more heavily doped materials where the pseudogap is absent. Our analysis suggests that $\chi_{Zn}(T)$ is complex and that the independent localized moment picture, even with the inclusion of the Kondo effect, does not explain the behavior over the entire experimental temperature range. Instead we conclude that the data is better described in terms of a Zn-induced re-distribution of quasiparticle spectral weight within a single band. Specifically the Curie-Weiss term induced by Zn doping reflects a narrow QP resonance close to $E_F$, with energy width ~ 15K, drawn predominantly from states in the band at or above the pseudogap energy. $\chi(T, y)$ for the OD samples points towards a gradual increase in the QP spectral weight with decreasing temperature, although here we cannot rule out the localized moment picture. The fundamentally different behavior for the UD and OD compounds is attributed to the presence of the PG in the former and to its absence in the latter, as found by other experimental studies [27-29].

## ACKNOWLEDGEMENTS

We thank J. L. Tallon for helpful comments and suggestions. SHN acknowledges financial support from the Commonwealth Scholarship Commission (U.K.), and the Department of Physics, Cambridge University. RSI acknowledges financial support from Trinity College, Cambridge, and the Cambridge Commonwealth Trust (U.K.). All experimental work in Cambridge was supported by the EPSRC (U.K.).

**References:**

*Corresponding author. E-mail: salehnaqib@yahoo.com[1]   J. M. Tarascon, L. H. Greene, P. Barboux, W. R. Mckinnon, G. W. Hull, T. P. Orlando, K. A. Delin, S. Foner, and E. J. NcNiff, Jr., Phys. Rev. B **36**, 8393 (1987).

[2]   T. R. Chien, Z. Z. Whang, and N. P. Ong, Phys. Rev. Lett. **67**, 2088 (1991).

[3]   S. Zagoulaev, P. Monod and J. Jegoudez, Phys. Rev. **B 52**, 10474 (1995).

[4]   J. L. Tallon, J. R. Cooper, P. S. I. P. N. de Silva, G. V. M. Williams, and J. W. Loram, Phys. Rev. Lett. **75**, 4114 (1995).

[5]   A. M. Finkel'stein, V. E. Kataev, E. F. Kukovitskii and G. B. Teitel'baum, Physica C **168**, 370 (1990).

[6]   H. Alloul, P. Mendels, H. Casalta, J. F. Marucco, and J. Arabski, Phys. Rev. Lett. **67**, 3140 (1991).

[7]   A. V. Mahajan, H. Alloul, G. Collin, and J. F. Marucco, Phys. Rev. Lett. **72**, 3100 (1994).

[8]   S. Zagoulaev, P. Monod, and J. Jegoudez, Physica C **259**, 271 (1996).

[9]   G. Xiao, M. S. Cieplak, J. Q. Xiao, and C. L. Chien, Phys. Rev. B **42**, 8752 (1990).

[10]  N. Ishikawa, N. Kuroda, H. Ikeda, R. Yoshizaki, Physica C **203**, 284 (1992).

[11]  S. H. Pan, E. W. Hudson, K. M. Lang, H. Eisaki, S. Uchida, and J. C. Davis, Nature **403**, 746 (2000).

[12]  J. W. Loram, K. A. Mirza, J. M. Wade, J. R. Cooper and W. Y. Liang, Physica C **235-240**, 134 (1994).

[13]  J. R. Cooper and J. W. Loram, J. Phys. I (France) **6**, 2237 (1996).

[14]  H. F. Fong, P. Bourges, Y. Sidis, L. P. Regnault, J. Bossy, A. Ivanov, D. L. Milius, I. A. Aksay and B. Keimer, Phys. Rev. Lett. **82**, 1939 (1999).

[15]  R. S. Islam, Ph.D. thesis, University of Cambridge (UK), 2005 (unpublished).

[16]  J. W. Loram, J. Luo, J.R. Cooper, W. Y. Liang, and J. L. Tallon, J. Phys. Chem. Solids **62**, 59 (2001).

[17]  R. S. Islam, J. R. Cooper, J. W. Loram, and S. H. Naqib, Physica C **460-462**, 753, (2007).
11

**Figure captions:**

Fig. 1. (Color online) $\chi(T)$ data of $La_{2-x}Sr_xCu_{1-y}Zn_yO_4$ compounds for the Sr and Zn contents (y) shown.

Fig. 2. (Color online) Representative plots of $\chi$ vs. Zn content (y) at fixed temperatures ($T_0$) for $La_{1.85}Sr_{0.15}Cu_{1-y}Zn_yO_4$. Values of $T_0$ are shown. Dashed straight lines show linear fits.

Fig. 3. (Color online) Plots of the fitted magnetic susceptibility, $\chi_{fit}(T) \equiv a_0 + a_1 y$, of $La_{2-x}Sr_xCu_{1-y}Zn_yO_4$ compounds for the Sr and Zn contents (y) shown. The crossing temperature (see text for details), $T_{cs}$, is marked in Figs. 3a and 3b.

Fig. 4. (Color online) Difference plots, giving the errors $\Delta\chi = [\chi(T) - \chi_{fit}(T)]$ for $La_{2-x}Sr_xCu_{1-y}Zn_yO_4$ compounds for the Sr and Zn contents (y) shown.

Fig. 5. (Color online) (a) Fitted values $a_0(T)$ (symbols) and raw data near $T_c$ (dashed lines) for the Zn-free samples for all $La_{2-x}Sr_xCu_{1-y}Zn_yO_4$ compounds studied. The Sr contents (x) are given in the figure. (b) the susceptibility increment $a_1(T)$ per %Zn. (c) Plots of $a_1(T)T$ for all $La_{2-x}Sr_xCu_{1-y}Zn_yO_4$ compounds. For $T < 100$ - 150K (depending on x), $a_1(T)T$ is nearly constant so the Zn induced magnetic contribution approximately follows a Curie law with Curie constant $C_{LT}$. (d) Plots of $a_1(T)$ vs. $1/T$ for all $La_{2-x}Sr_xCu_{1-y}Zn_yO_4$ compounds. Dashed lines show fits used to obtain the slopes ($C_{HT}$) and intercepts $\Delta\chi_{0,HT}$ at $1/T = 0$.

Fig. 6. (Color online) Left scale, low and high temperature estimates of the Curie constant $C_{LT}$ (triangles) and $C_{HT}$ (closed circles) and right scale, the "dilution" term $\Delta\chi_{0,HT}$ (squares) vs. Sr content (x), deduced from fits to the two component Eq. 2. The dilution term at low temperature (not shown) is $\Delta\chi_{0LT} = 0 \pm 0.007 \times 10^{-4}$ emu/mol%Zn. The quantity $10^{-2}\chi(400K)$ is also plotted vs. Sr content (diamonds, right scale) and gives an estimate of the spin-dilution effect expected for 1% Zn substitution (see text).

Fig. 7. (Color online) (a) $a_0^* = d(a_0T)/dT$ for the Zn-free material (smoothed over 7 data points) and (b) $a_1^* = d(a_1T)/dT$ for the increment per %Zn (smoothed over 13 data points). These plots narrow the Curie-Weiss term and show more clearly higher temperature changes in the DOS.

Fig. 8. (Color online) Main: $a_1^*$ for 15% Sr. Above 60K the curve is derived from a linear fit for all Zn contents (shown in Fig. 7b). Below 60K curve shown for weakly or non-superconducting 2.0% and 2.4% Zn samples assuming a linear extrapolation of the Zn-free normal state $\chi^*$ to below $T_c$ (see text). Inset: $\chi^* = d(\chi T)/dT$ for 15% Sr containing 0% and 2.4% Zn. Fig. 8 suggests that the low temperature peak in Zn doped samples results from the re-distribution (softening) of higher energy states in the Zn-free band.

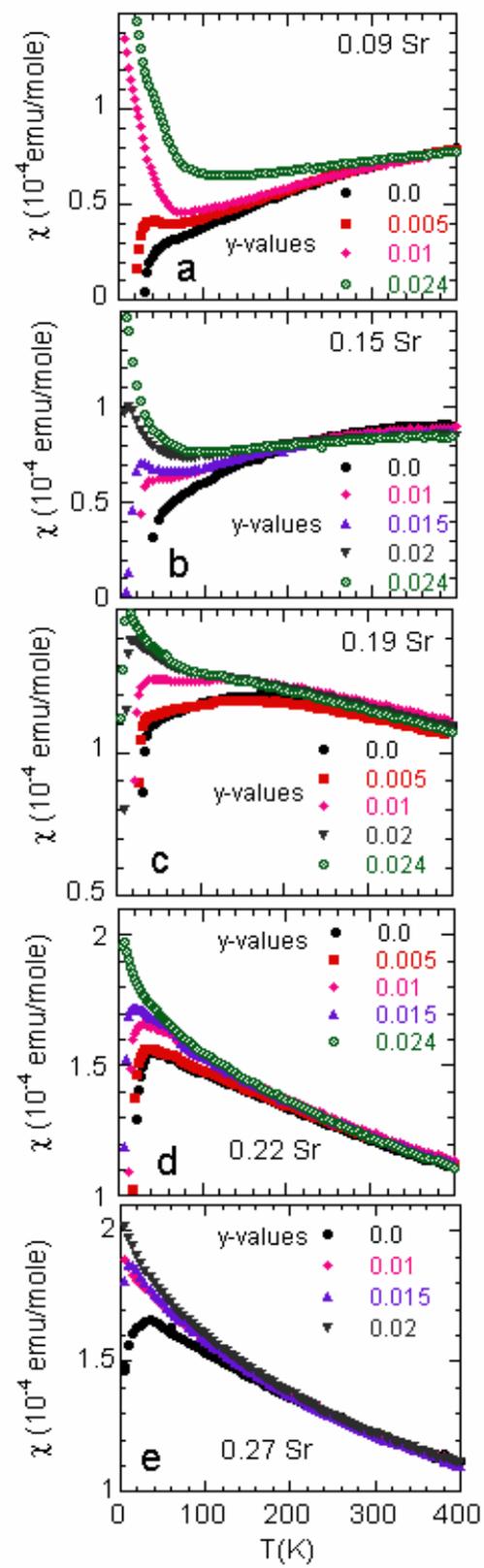

**Figure 1**

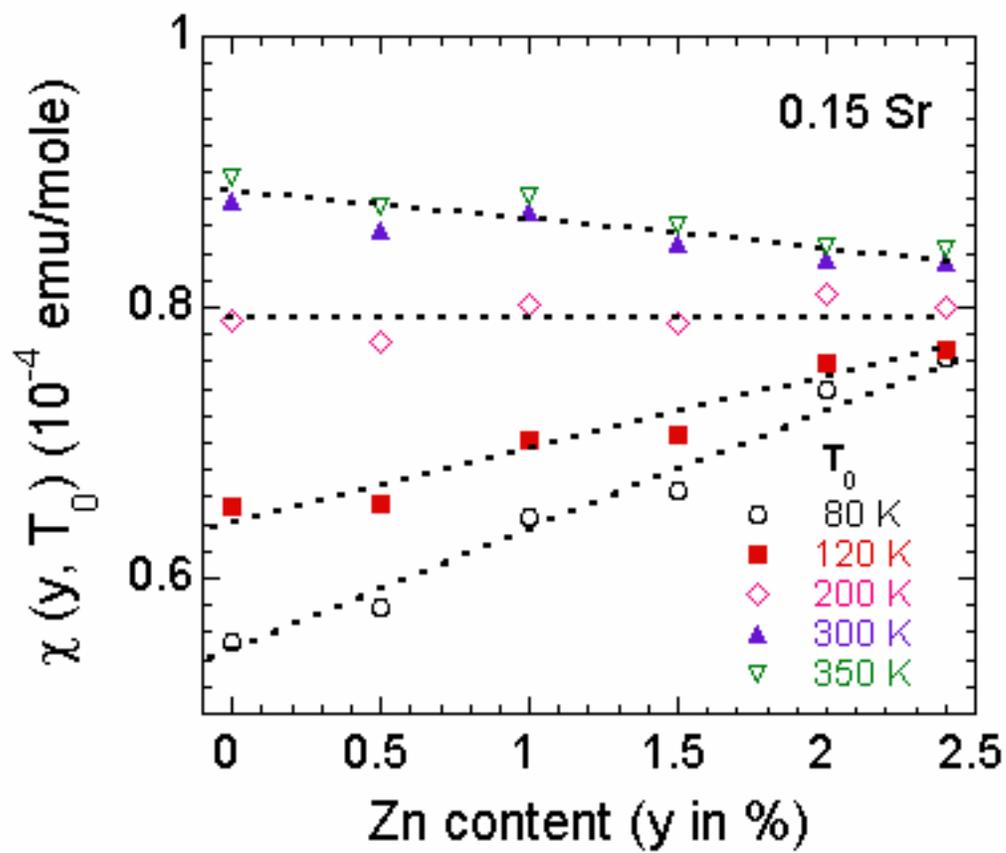

**Figure 2**

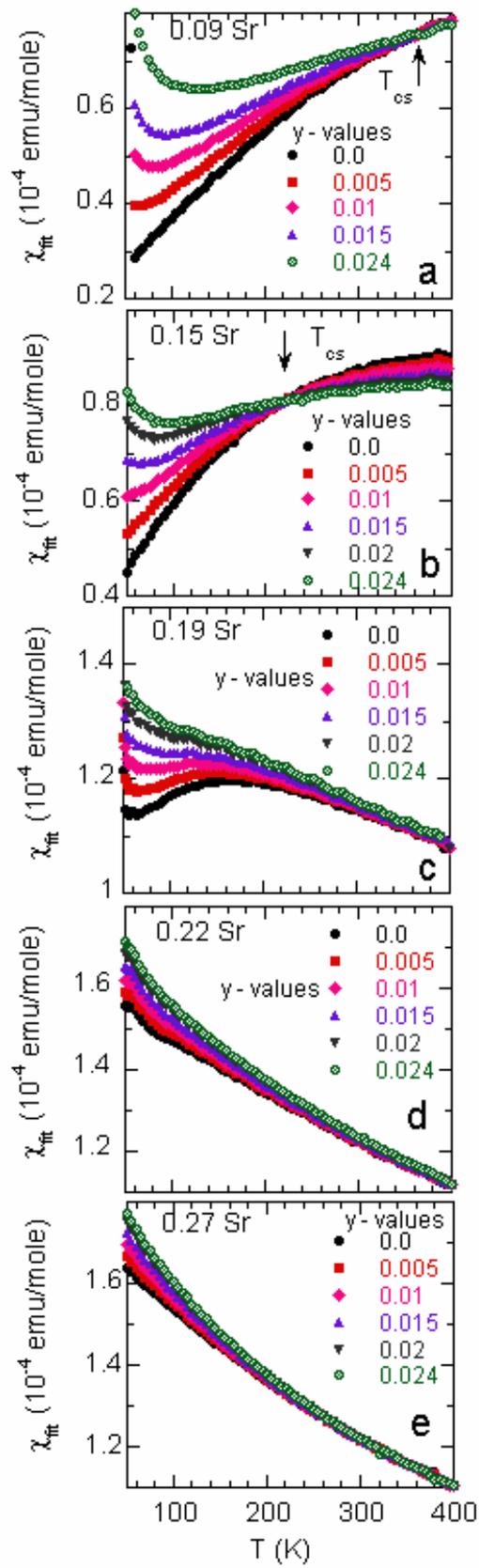

**Figure 3**

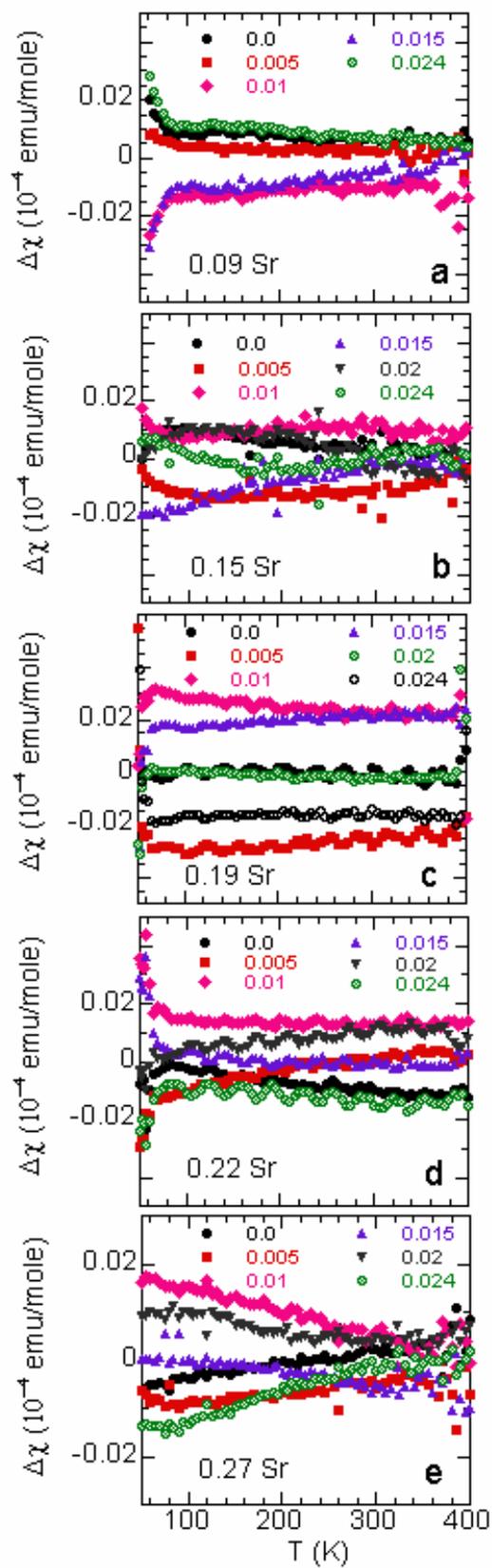

**Figure 4**

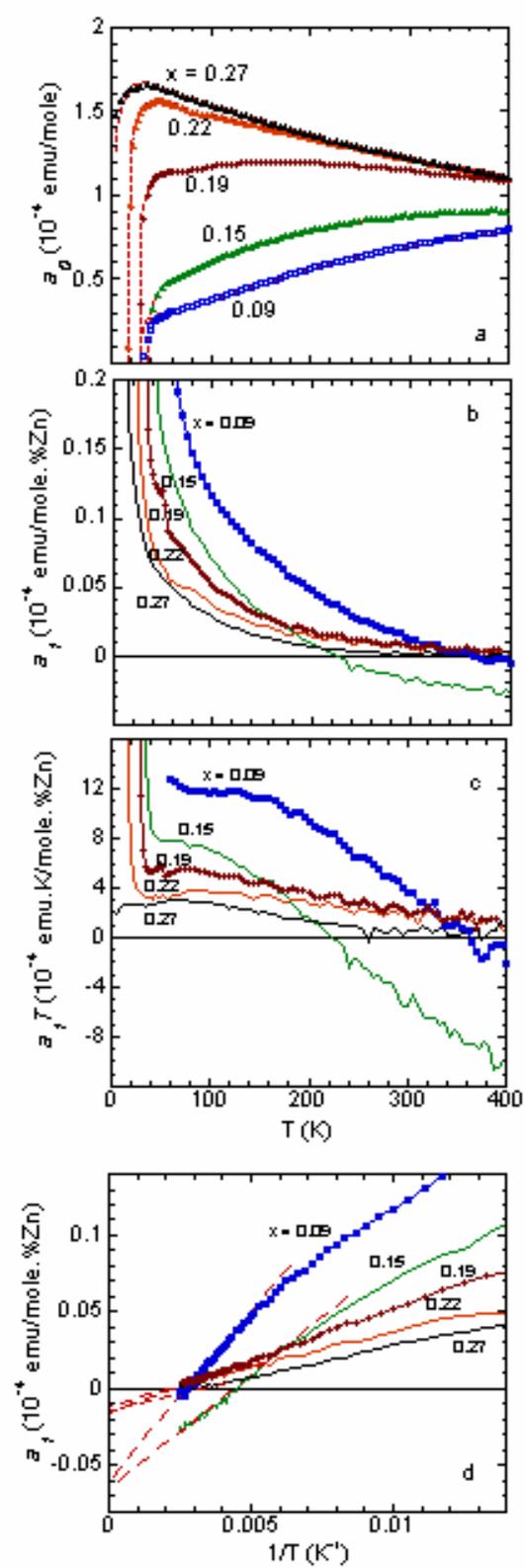

**Figure 5**

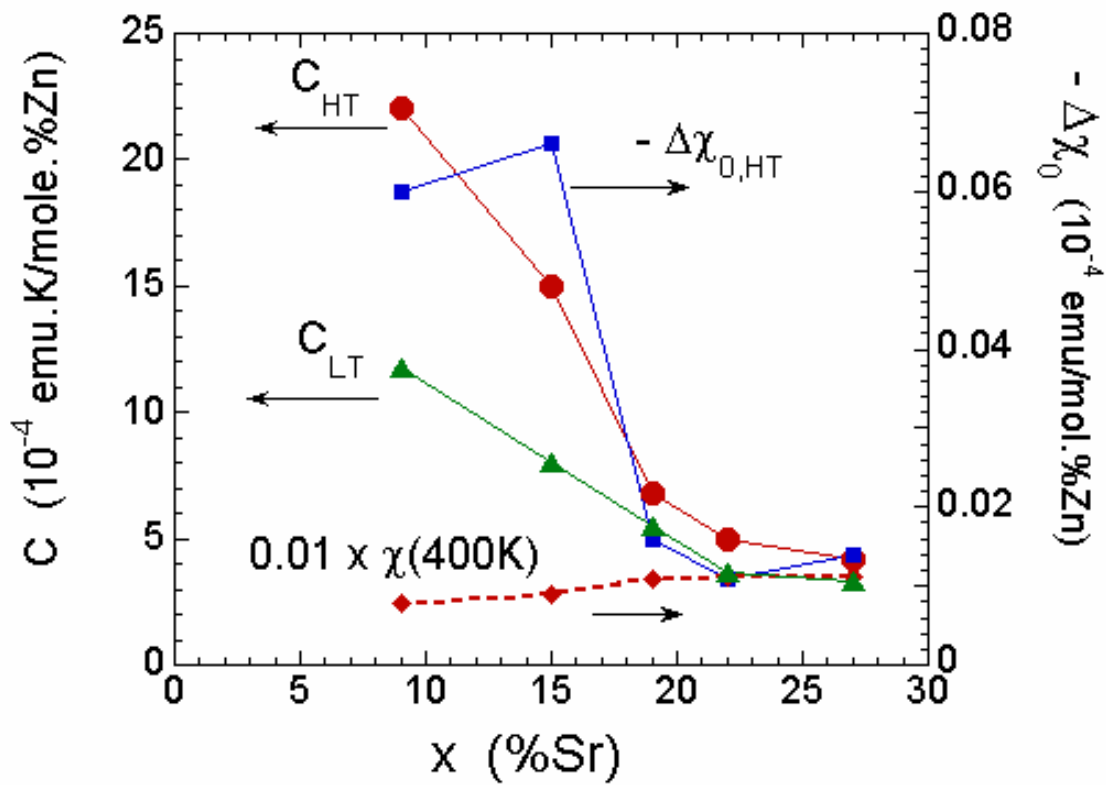

**Figure 6**

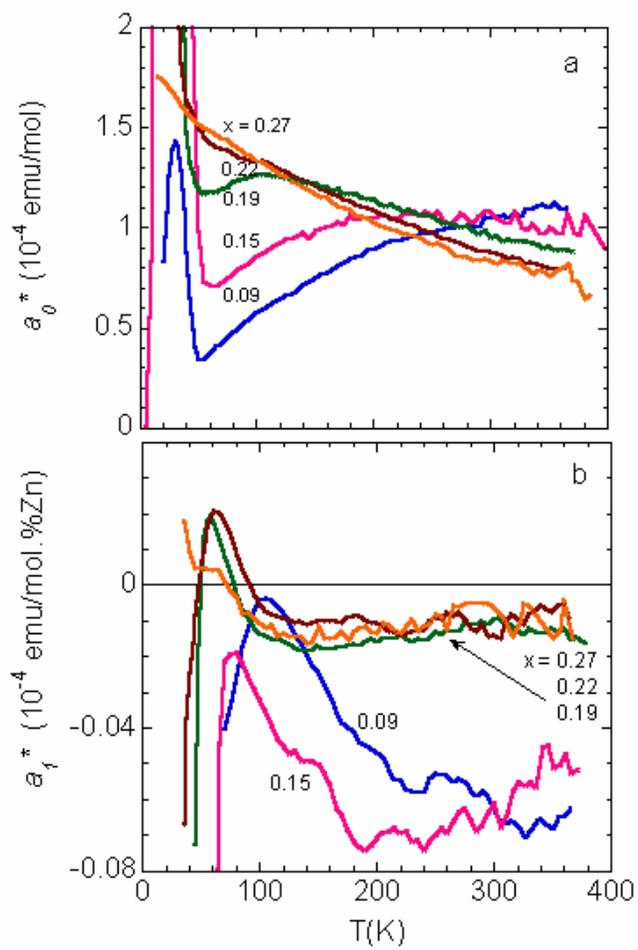

**Figure 7**

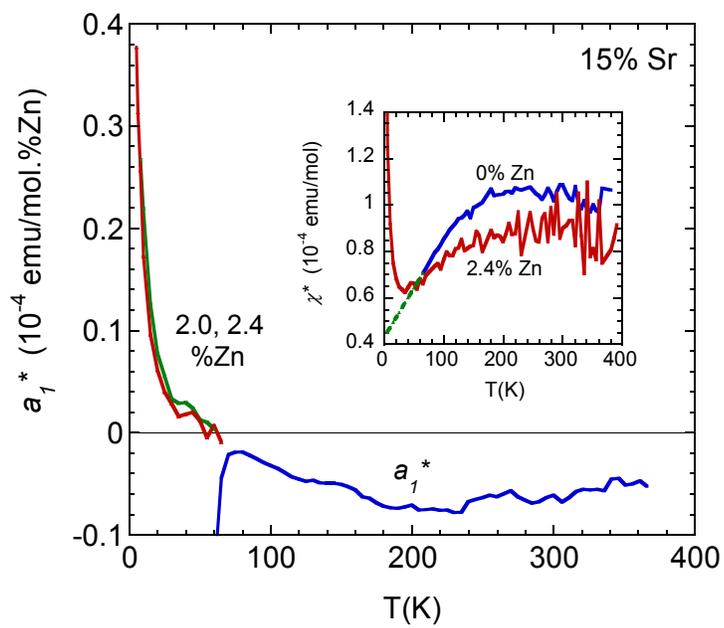

**Figure 8**